# Ruppeiner thermodynamic geometry, microstructure, quasinormal modes and greybody factors of the Einstein–Skyrme black hole

Shakhboz Khasanov[1,*], Akmal Safarov[1], Bo Yang[2],
Habtamu Menberu Tedila[3]

[1]Institute of Nuclear Technologies, Samarkand State University named after Sharof Rashidov, Samarkand 140104, Uzbekistan
[2]Institute of Modern Physics, Chinese Academy of Sciences, Lanzhou 730000, China
[3]McGill University, 3550 University Street, Montreal, QC H3A 2A7, Canada

*Email: shakhbozkhasanov93@gmail.com

**Abstract** We investigate the thermodynamic microstructure, quasinormal mode spectrum, and greybody factors of the exact, static, spherically symmetric black hole in Einstein–Skyrme theory, building on the recently established first law that promotes the Skyrme couplings $K$ and $\lambda$ to extensive variables. Using Ruppeiner geometry, we construct the scalar curvature from the Hessian of the mass. The curvature remains finite at the second-order (heat-capacity) phase transition but diverges in the extremal, zero-temperature limit. Throughout the physically admissible region the curvature retains a single sign, indicating microscopic interactions of one dominant (attractive) type whose strength grows toward extremality. We perform a Joule–Thomson-like isenthalpic expansion and prove analytically that the corresponding coefficient is strictly negative across the entire physical parameter space, implying the black hole always cools as $\lambda$ increases at fixed mass, with no inversion temperature. Turning to perturbations, we derive the effective potential for massless scalar fields. The solid-angle-deficit coupling $K$ controls its shape—substantially lowering the barrier—while at fixed $K$ the quartic coupling $\lambda$ produces only minor changes. Rigorous lower bounds on greybody factors are obtained in closed form using Visser's method: increasing $K$ raises the bound, since a larger horizon lowers the centrifugal barrier, whereas increasing $\lambda$ mildly suppresses low-frequency transmission. Our results provide the first comprehensive study of the thermodynamic microstructure and perturbative spectroscopy of this rare analytic hairy black hole, complementing and extending the existing thermodynamic analysis into previously unexplored territory.

## 1 Introduction

Black holes serve as remarkable theoretical laboratories in which the fundamental principles of general relativity, quantum field theory, and statistical mechanics converge. The celebrated laws of black hole thermodynamics, formulated by Bekenstein [1,2] and Hawking [3,4], established a profound correspondence between the macroscopic properties of event horizons—such as area, surface gravity, and mass—and the familiar thermodynamic concepts of entropy, temperature, and energy. These laws have since provided deep insights into the quantum nature of gravity and continue to guide efforts toward a complete theory of quantum spacetime [5–7].

A major conceptual advance of the past decade has been the reinterpretation of the cosmological constant $\Lambda$ as a thermodynamic pressure, leading to the paradigm of black hole chemistry in an extended phase space [8–10]. Within this framework, the black hole mass is naturally identified with enthalpy rather than internal energy, and the thermodynamic phase space is enriched by a pressure–volume sector. This perspective has revealed a striking analogy between the phase behavior of charged anti-de Sitter black holes and that of Van der Waals fluids, including first-order phase transitions, critical phenomena, and swallow-tail structures in the Gibbs free energy [11–14]. These developments have inspired numerous extensions, encompassing superfluid-like behavior [15], reentrant phase transitions [16], triple points [17], and holographic interpretations of black hole degrees of freedom [18,19].

Despite this progress, a fundamental question persists: what constitutes the microscopic degrees of freedom of a black hole? In the absence of a complete theory of quantum gravity, one may turn to thermodynamic geometry for indirect guidance. The Ruppeiner geometry [20,21], which constructs a Riemannian metric on the thermodynamic state space from the Hessian of the entropy, provides precisely such a tool. The scalar curvature of the Ruppeiner metric encodes information about thermodynamic fluctuations and has been interpreted as a measure of the type and strength of interactions among the microstructural constituents of a thermodynamic system [22–24]. For ordinary fluids, positive (negative) Ruppeiner curvature corresponds to repulsive (attractive) interactions, while a vanishing curvature characterizes the ideal gas. Applied to black holes, this approach has revealed intriguing features: charged anti-de Sitter black holes exhibit both attractive and repulsive microstructural interactions depending on the thermodynamic branch [25–27], and the Ruppeiner curvature diverges at phase transitions with universal critical exponents matching those of the Van der Waals fluid [25].

In parallel, the study of black hole perturbations—specifically quasinormal modes (QNMs)—provides a complementary probe of black hole spacetimes through their characteristic oscillation frequencies [28–30]. The quasinormal mode spectrum depends solely on the background geometry and carries a unique signature of the black hole parameters, making QNMs natural observables for gravitational-wave astronomy. Closely related are the greybody factors, which quantify the modification of Hawking radiation by the curved-spacetime potential barrier surrounding the event horizon [31,32]. Together, QNMs and greybody factors offer a perturbative spectroscopy of the near-horizon geometry that is sensitive to the matter content sourcing the black hole.

Among matter models coupled to gravity, the Skyrme model occupies a distinguished position as a nonlinear field theory that supports topologically nontrivial soliton solutions—Skyrmions—identified with baryons in the low-energy limit of quantum chromodynamics [33,34]. When coupled to Einstein gravity, the resulting Einstein–Skyrme system admits self-gravitating Skyrmion and black hole solutions [35–38]. Owing to the strong nonlinearity and higher-derivative structure of the Skyrme equations, exact analytic solutions are exceedingly rare; most known configurations have been obtained numerically. A notable exception is the exact, static, spherically symmetric black hole found by Canfora and Maeda [35] via a generalized hedgehog ansatz. This solution is characterized by two independent coupling parameters: $K$, governing the strength of the quadratic sigma-model term, and $\lambda$, controlling the quartic Skyrme interaction. Their respective contributions to the metric produce a solid-angle-deficit-like deformation [39] and an effective charge-like term.

The thermodynamics of Einstein–Skyrme black holes has been studied previously. Flores-Alfonso and Quevedo [40] examined the extended thermodynamics of self-gravitating Skyrmions with a cosmological constant, interpreting $\Lambda$ as thermodynamic pressure. Khan, Ganai, and Upadhyay [41] investigated quantum-corrected thermodynamics and P–V criticality of the same system. Very recently, Senjaya and Klaypong [42] formulated a novel extended first law of black hole thermodynamics for the non-AdS Canfora–Maeda black hole by promoting $K$ and $\lambda$ to extensive thermodynamic variables, and demonstrated that the resulting phase structure exhibits second-order phase transitions signaled by heat capacity divergences and continuous Gibbs free energy behavior.

However, several fundamental questions raised by these developments remain entirely open. First, no investigation of the thermodynamic microstructure of the Einstein–Skyrme black hole via Ruppeiner geometry has been undertaken. It is unknown whether the microscopic interactions in this system are dominantly attractive or repulsive, how the Skyrme couplings influence the nature of these interactions, and whether the Ruppeiner curvature exhibits the universal divergence at the phase transition found in Ref. [42].

Second, no study of quasinormal modes or greybody factors for the exact Canfora–Maeda black hole exists in the literature. The impact of the Skyrme parameters K and λ on the perturbative spectrum and Hawking radiation transmission remains completely unexplored. Third, the Joule–Thomson-like expansion and the microstructure phase diagram—important thermodynamic tools extensively employed for charged and rotating AdS black holes [43–45]—have never been analyzed for the Einstein–Skyrme system.

The present work aims to fill these gaps comprehensively. We conduct the first analysis of the Ruppeiner thermodynamic geometry and microstructure of the exact Einstein–Skyrme black hole within the extended thermodynamic framework of Senjaya and Klaypong [42]. We compute the normalized Ruppeiner scalar curvature and analyze its sign, divergences, and dependence on the Skyrme couplings, elucidating the nature of microscopic interactions and their connection to phase transitions. We further perform a Joule–Thomson-like isenthalpic expansion analysis and construct a microstructure phase diagram in the entropy–coupling plane. Turning to the perturbative sector, we derive the effective potential for massless scalar field perturbations and establish rigorous lower bounds on the greybody factors using Visser's general method [46]. Throughout, we examine how the Skyrme parameters K and λ modify the physical observables relative to the Schwarzschild baseline.

This paper is organized as follows. In Sect. 2, we briefly review the exact Einstein–Skyrme black hole solution and its thermodynamic framework. Section 3 develops the Ruppeiner thermodynamic geometry and analyzes the microstructure. In Sect. 4, we construct the microstructure phase diagram and analyze the isenthalpic expansion. Section 5 is devoted to scalar field perturbations, the effective potential, and greybody factor bounds. Finally, Sect. 6 summarizes our results and discusses their implications.

## 2 Black hole solution and thermodynamic framework

In this section, we briefly review the exact static and spherically symmetric black hole solution in the Einstein–Skyrme theory and the associated thermodynamic framework established in [42]. The presentation is self-contained and introduces the notation used throughout the remainder of this work.

### 2.1 The Canfora–Maeda solution

The Einstein–Skyrme system is governed by the action $S = S^G + S_{skm}$, where $S^G$ denotes the Einstein–Hilbert action and the Skyrme sector describes an SU(2)-valued scalar field $U(x^U)$ with dynamics determined by the quadratic sigma-model coupling $K = F\pi^2/4$ and the quartic Skyrme coupling $\lambda = 4e^2F\pi^2$, where $F\pi$ is the pion decay constant and e the Skyrme parameter [33–35]. The coupled Einstein–Skyrme equations admit the exact, static, spherically symmetric black hole solution [35]:

$$ds^2 = -f(r)dt^2 + \frac{dr^2}{f(r)} + r^2(d\vartheta^2 + sin^2\vartheta\, d\varphi^2) \tag{1}$$

with the metric function:

$$f(r) = 1 - 8\pi GK - \frac{2GM}{r} + \frac{4\pi GK\lambda}{r^2} \tag{2}$$

The parameter K controls the solid-angle-deficit-like deformation, yielding the asymptotic behavior $f(r \to \infty) \to 1 - 8\pi GK$ characteristic of a Barriola–Vilenkin-type geometry [39], while λ generates an effective charge contribution via the $r^{-2}$ term. The event horizon $r_h$ is defined as the largest positive root of $f(r_H) = 0$.

Two structural features of this background matter for the perturbative analysis below. First, because $f(r \to \infty) \to 1 - 8\pi GK < 1$, the spacetime is not asymptotically flat but carries a solid-angle deficit of Barriola–Vilenkin type; a global rescaling $\tilde{r} = r\sqrt{(1-8\pi GK)}$, $\tilde{t} = t/\sqrt{(1-8\pi GK)}$ restores asymptotic flatness at the cost of a corresponding rescaling of the frequency, which must be kept in mind when comparing spectra at different K. Second, the $+4\pi GK\lambda/r^2$ term enters f(r) with the same positive sign as the charge term $Q^2/r^2$ of a Reissner–Nordström black hole, so the quartic Skyrme coupling behaves as an effective electric-charge-like contribution. Throughout, we set G = 1 in all numerical evaluations and figures.

### 2.2 Extended first law and thermodynamic quantities

Following [42], we treat the Skyrme couplings K and λ as extensive thermodynamic variables with conjugate potentials $\Phi_K$ and $\Phi_\lambda$. The extended first law takes the form: We note that the potential Φ_K = 2√π(πλ − 2GS)/√(GS) changes sign at πλ = 2GS, reflecting the competition between the λ- and entropy-dependent contributions to ∂M/∂K; this is a smooth change of sign of a conjugate potential and does not by itself signal an instability.

$$dM = TdS + \Phi_K dK + \Phi_\lambda d\lambda \tag{3}$$

The mass expressed as a function of the extensive variables S, K, and λ reads:

$$M(S,K,\lambda) = \frac{S(1-8\pi GK)+4\pi^2 K\lambda}{2\sqrt{\pi GS}} \tag{4}$$

Partial differentiation yields the Hawking temperature:

$$T = \frac{S(1-8\pi GK)-4\pi^2 K\lambda}{4\sqrt{\pi G S^{\frac{3}{2}}}} \tag{5}$$

the conjugate potentials:

$$\Phi_K = \frac{2\sqrt{\pi}(\pi\lambda - 2GS)}{\sqrt{GS}}, \quad \Phi_\lambda = \frac{2K\pi^{3/2}}{\sqrt{GS}} \tag{6}$$

and the heat capacity at fixed Skyrme couplings:

$$C_H = \frac{2S[S(1-8\pi GK)-4\pi^2 K\lambda]}{S(8\pi GK-1)+12\pi^2 K\lambda} \tag{7}$$

The Gibbs free energy is:

$$G = M - TS = \frac{S(1-8\pi GK)+12\pi^2 K\lambda}{4\sqrt{\pi GS}} \tag{8}$$

The heat capacity diverges at the transition entropy:

$$S_t = \frac{12\pi^2 K\lambda}{1-8\pi GK} \tag{9}$$

signaling a second-order phase transition in the Ehrenfest classification [42].

## 3 Ruppeiner thermodynamic geometry and microstructure

The Ruppeiner geometry provides a powerful framework for probing the microstructure of thermodynamic systems from their macroscopic observables [20–25]. In this section, we construct the Ruppeiner geometry for the Einstein–Skyrme black hole and analyze the nature of microscopic interactions encoded in the thermodynamic curvature scalar.

### 3.1 Construction of the Ruppeiner metric

The Ruppeiner metric is defined on the space of equilibrium states by the Hessian of the entropy with respect to the extensive variables. For a system with internal energy M and extensive variables (S, K, λ), the line element of the Ruppeiner geometry in the energy representation is given by:

$$dl^2 = \frac{1}{T}\frac{\partial^2 M}{\partial x^i \partial x^j} dx^i dx^j \tag{10}$$

Because the mass is linear in each Skyrme coupling, the diagonal second derivatives $\partial^2 M/\partial\lambda^2$ and $\partial^2 M/\partial K^2$ vanish identically. This does not make the two-dimensional metric degenerate: the off-diagonal component $\frac{\partial^2 M}{\partial S\,\partial\lambda}$ (equivalently $\frac{\partial^2 M}{\partial S\,\partial K}$) is non-zero, so $\det(g) = -\frac{1}{T^2}\left(\frac{\partial^2 M}{\partial S\,\partial\lambda}\right)^2 \neq 0$ and the metric is regular. We therefore evaluate the genuine scalar curvature of the two-dimensional (S, λ) Ruppeiner metric at fixed K directly, without resorting to a one-dimensional reduction. The calculation yields the compact closed form:

$$R = \frac{1-8\pi GK}{4\pi^2 K\lambda - S(1-8\pi GK)} \tag{11}$$

Here we have adopted the energy-representation metric $g_{ij} = \frac{1}{T}\frac{\partial^2 M}{\partial x^i\,\partial x^j}$ with $x = (S, \lambda)$; the overall sign convention is discussed below. We emphasize that R in Eq. (11) is the genuine two-dimensional thermodynamic curvature, not a heat-capacity-based proxy, and that its attractive-versus-repulsive interpretation is convention dependent and is therefore stated below in convention-independent terms.

## 3.2 Computation of the thermodynamic curvature

Introducing the extremal entropy $S_{ext} = \frac{4\pi^2 K\lambda}{1-8\pi GK}$, the curvature (11) takes the remarkably simple form $R = \frac{1}{S_{ext}-S}$. Three features follow at once. First, R diverges as $S \to S_{ext}$ from above, that is, in the extremal limit T → 0 where the small–black-hole branch terminates, rather than at the heat-capacity transition. Second, at the transition entropy $S_t = \frac{12\pi^2 K\lambda}{1-8\pi GK} = 3S_{ext}$ the curvature is finite, $R(S_t) = -\frac{1-8\pi GK}{8\pi^2 K\lambda}$; the second-order transition diagnosed by the divergence of $C_H$ is thus not accompanied by a curvature singularity, since it corresponds to $\frac{\partial^2 M}{\partial S^2} = 0$, where the metric remains regular. Third, throughout the physical region $S > S_{ext}$ (where T > 0) the curvature keeps a single sign and |R| decreases monotonically with S, vanishing for large black holes. Table 1 summarizes this behavior.

**Table 1** Summary of Ruppeiner curvature behavior in different thermodynamic regimes

| Regime | Entropy range | Sign of R | Behaviour |
|---|---|---|---|
| Extremal limit | $S \to S_{ext}$ | $R \to -\infty$ | Divergence (T→0) |
| Small BH | $S_{ext} < S < S_t$ | $R < 0$ | Attractive-type |
| Transition | $S = S_t = 3S_{ext}$ | finite, $R < 0$ | No divergence |
| Large BH | $S \to \infty$ | $R \to 0^-$ | Non-interacting |

## 3.3 Microstructure analysis: dependence on Skyrme couplings

Figure 1 displays the Ruppeiner scalar curvature R as a function of the entropy S for fixed K = 0.015 and several values of the quartic Skyrme coupling λ. Three features emerge. First, each curve diverges at the extremal entropy $S_{ext}(\lambda) = \frac{4\pi^2 K\lambda}{1-8\pi GK}$, which shifts to larger entropies as λ increases; this is the zero-temperature limit in which the horizon becomes degenerate, not the heat-capacity phase transition of Ref. [42]. Second, R is negative throughout the physical region $S > S_{ext}$ and grows in magnitude as extremality is approached, indicating microscopic interactions of a single dominant character that strengthens toward the extremal limit. Third, |R| decreases monotonically with increasing S, so that large black holes approach a non-interacting, asymptotically flat-geometry regime. We emphasize that the sign of R is convention dependent—it reverses under an overall sign redefinition of the thermodynamic metric—so the robust, convention-independent statements are the location of the divergence (extremality), the regularity at $S_t$, and the absence of any sign change within the physical domain. We stress that this absence of a sign change is not a mere caveat of the construction but a finding in its own right. It differs qualitatively from the Wei–Liu–Mann picture familiar from charged AdS black holes [19, 25], in which the Ruppeiner curvature changes sign along the small–black-hole branch and thereby signals a crossover between repulsive and attractive microstructure regimes. For the Einstein–Skyrme black hole no such repulsive–attractive crossover occurs: the effective microscopic interactions retain a single dominant character over the entire physical domain.

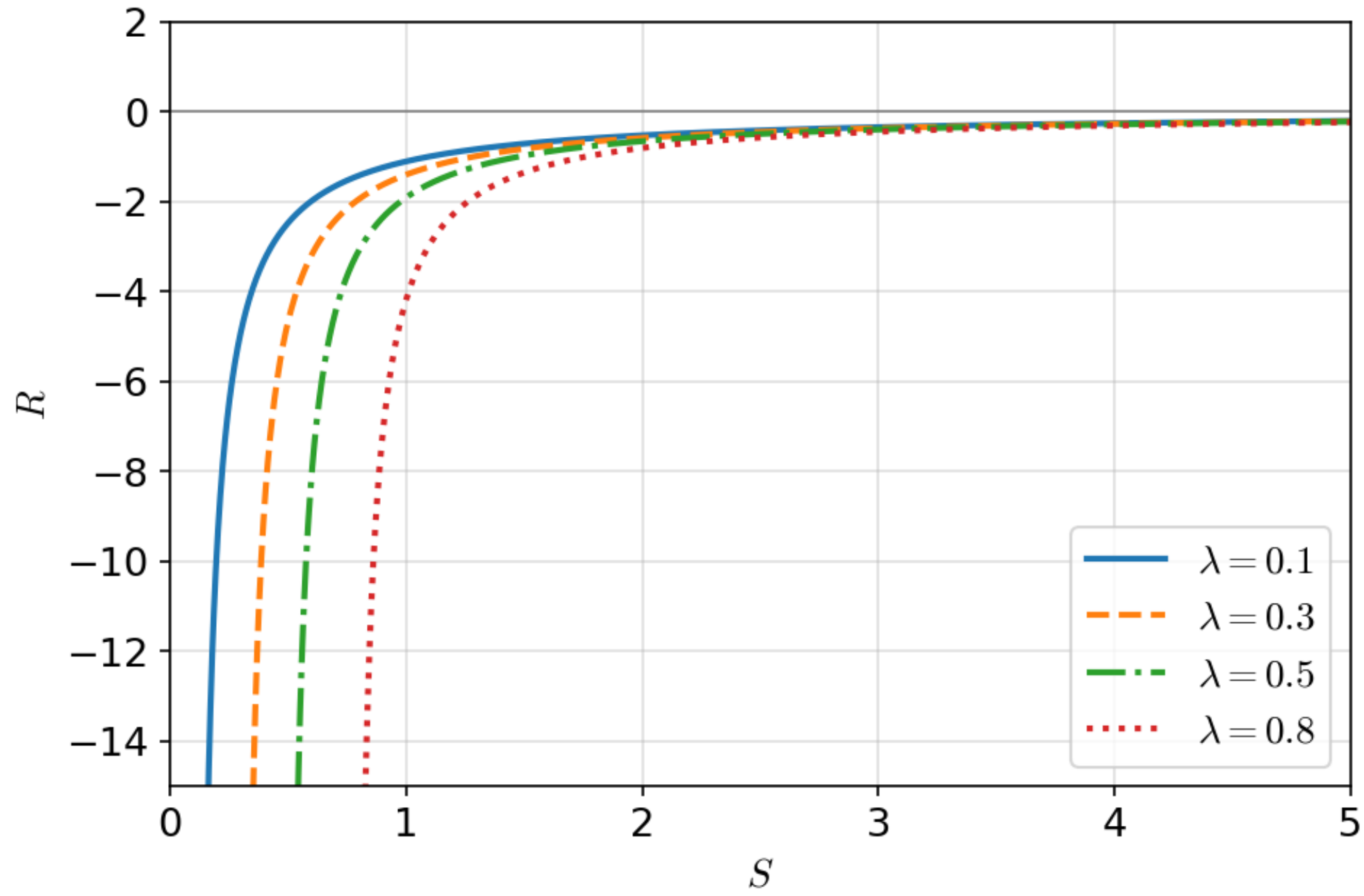

**Fig. 1** Ruppeiner curvature R as a function of entropy S for various values of the quartic Skyrme coupling λ, with K = 0.015 and G = 1. Each curve diverges at the extremal entropy $S_{ext}(\lambda)$ $(T \to 0)$ and remains negative throughout the physical region $S > S_{ext}$.

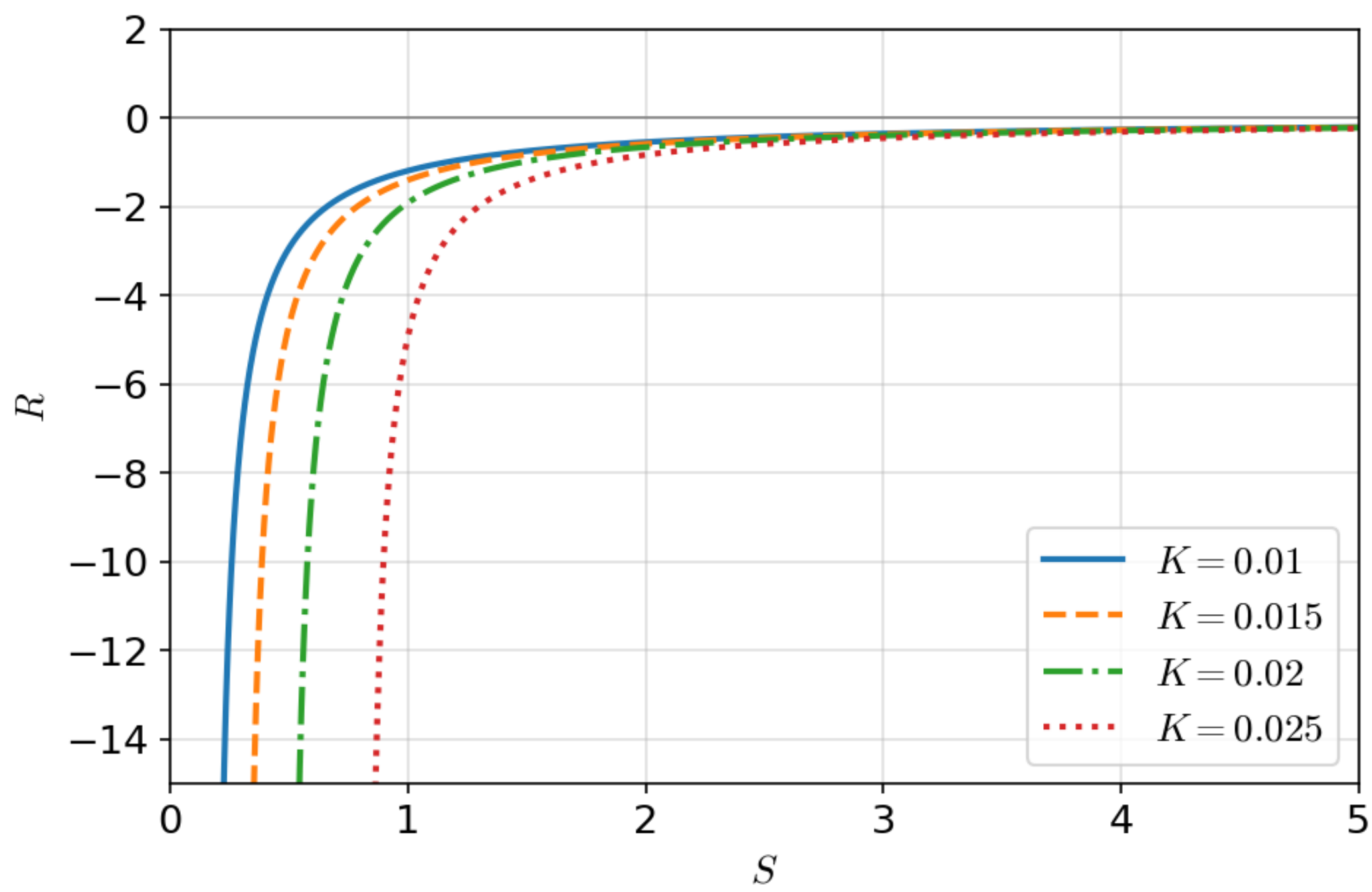

**Fig. 2** Ruppeiner curvature R as a function of entropy S for various values of K, with λ = 0.3 and G = 1. Increasing K shifts the extremal divergence $S_{ext}$ to larger entropies.

Figure 2 shows the corresponding behavior when the quadratic coupling K is varied at fixed λ = 0.3. Each curve again diverges at the extremal entropy $S_{ext}(K) = \frac{4\pi^2 K\lambda}{1-8\pi GK}$, whose location shifts to larger entropy as K grows. The same monotonic shift applies to the heat-capacity transition $S_t = 3S_{ext}$: explicitly, $\frac{dS_t}{dK} = \frac{12\pi^2\lambda}{(1-8\pi GK)^2} > 0$ throughout the physical range K < 1/(8πG), so both the extremal point and the transition move to larger entropy as K increases. As in Fig. 1, R is negative throughout the physical region and |R| decreases monotonically with S.

### 3.4 Temperature dependence of the Ruppeiner curvature

Figures 3 and 4 present the Ruppeiner curvature in the R–T plane, obtained by eliminating the entropy in favor of the Hawking temperature along the physical branch. Because T vanishes both at the extremal point $S = S_{ext}$ and in the large-entropy limit S → ∞, the temperature is a non-monotonic function of S; the figures show the small–black-hole branch $S_{ext} < S < S(T_{max})$, along which T rises from zero. On this branch the curvature diverges as $T \to 0^+$, i.e. in the extremal limit, and stays finite at the temperature associated with the heat-capacity transition. This confirms, from the R–T perspective, that the curvature singularity is tied to extremality rather than to the second-order transition.

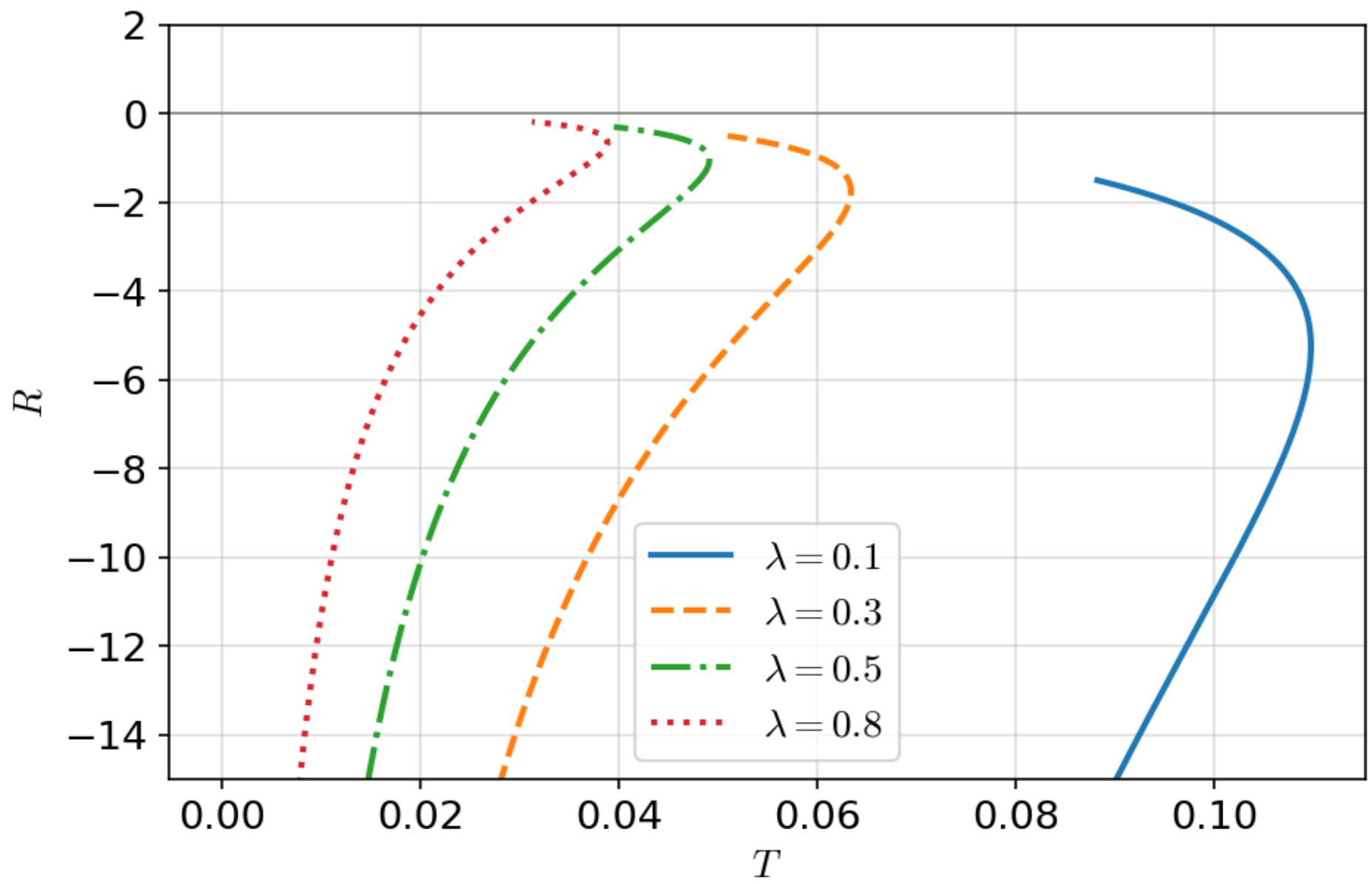

**Fig. 3** Ruppeiner curvature R as a function of temperature T for various λ at fixed K = 0.015. The divergence occurs as T → 0, the extremal limit.

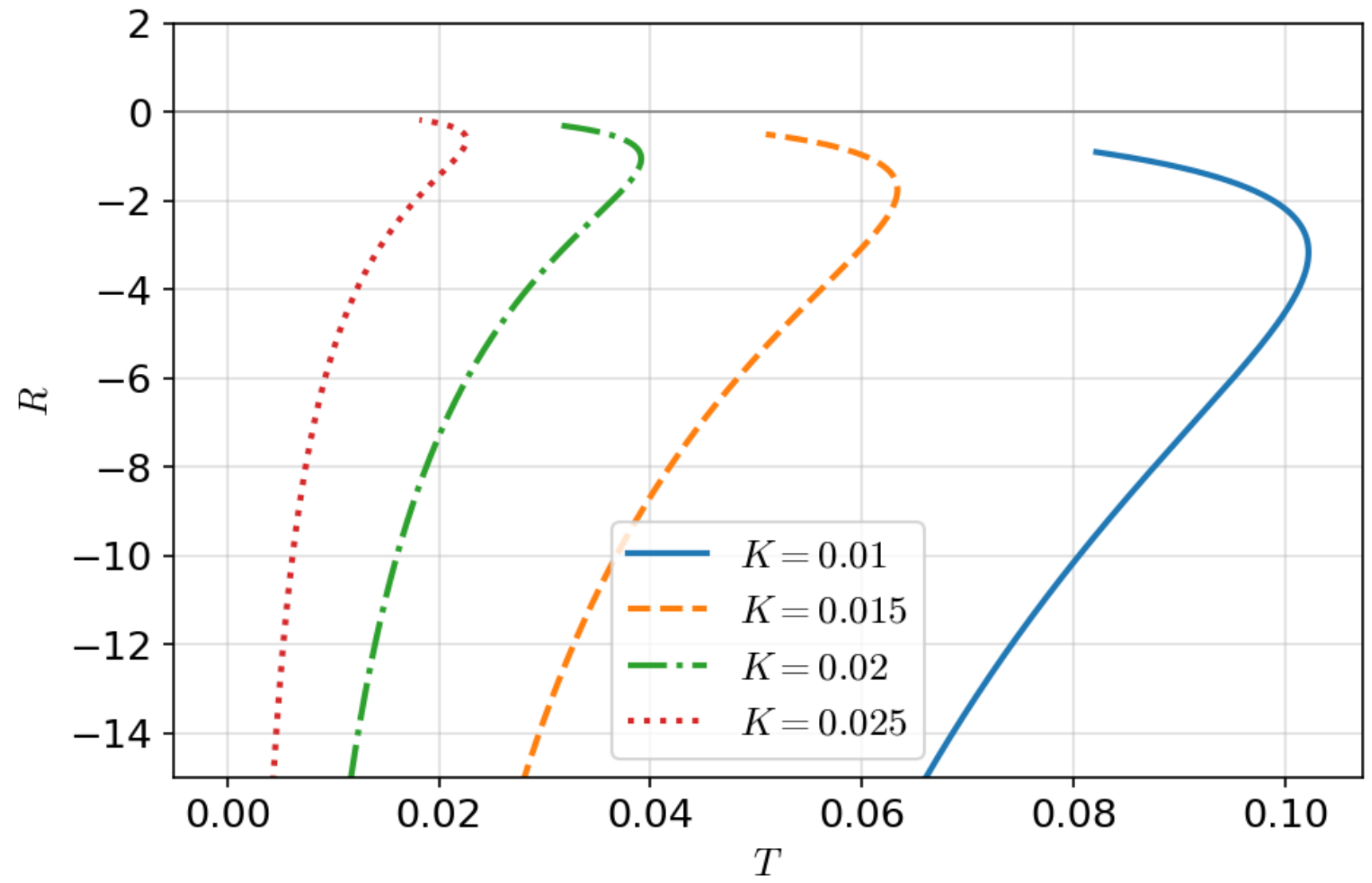

**Fig. 4** Ruppeiner curvature R as a function of temperature T for various K at fixed λ = 0.3.

# 4 Microstructure phase diagram and isenthalpic expansion

Since the curvature does not change sign within the physical region, the parameter space is organized more naturally by its stability and temperature structure than by a repulsive–attractive boundary. To visualize this, we construct a diagram in the (S, λ) plane at fixed K. The extremal line $S_{ext}(\lambda) = \frac{4\pi^2 K\lambda}{1-8\pi GK}$ bounds the physical region from below (T > 0 requires $S > S_{ext}$), while the transition entropy $S_t(\lambda) = 3S_{ext}$ separates the stable small–black-hole branch ($C_H > 0$) from the unstable large–black-hole branch ($C_H < 0$). We also study a Joule–Thomson-like isenthalpic expansion. In the present framework the quartic coupling λ enters the first law through the work term $\Phi_\lambda\, d\lambda$ and plays the role of a pressure-like variable; we therefore examine how the temperature responds to an isenthalpic (fixed-M) variation of λ. In the ordinary Joule–Thomson effect the coefficient $\mu = \left(\frac{\partial T}{\partial P}\right)_H$ controls whether a fluid cools or heats upon expansion. In the extended framework the mass M plays the role of the enthalpy, so the isenthalpic process is one at fixed M; no spatial expansion is implied, and the analogy rests solely on the $\Phi_\lambda\, d\lambda$ work term, with λ acting as the pressure-like variable.

## 4.1 Phase diagram and isenthalpic curves

The left panel of Fig. 6 displays this diagram for K = 0.015. The extremal line $S_{ext}$ (dashed) and the heat-capacity transition line $S_t = 3S_{ext}$ (solid) partition the (S, λ) plane into the unphysical region ($S < S_{ext}, T < 0$), the stable small–black-hole branch ($S_{ext} < S < S_t, C_H > 0$), and the unstable large–black-hole branch ($S > S_t, C_H < 0$). The right panel shows the transition temperature $T_t$ as a function of λ for several values of K. Using λ as the pressure-like variable, we define the Joule–Thomson-like coefficient as:

$$\mu_{JT} = \left(\frac{\partial T}{\partial \lambda}\right)_M \tag{12}$$

Evaluating this derivative at constant mass M using the chain rule and the first law (3), we obtain:

$$\mu_{JT} = \left(\frac{\partial T}{\partial S}\right)\left(\frac{\partial S}{\partial \lambda}\right)_M + \left(\frac{\partial T}{\partial \lambda}\right)_S \tag{13}$$

where $\left(\frac{\partial S}{\partial \lambda}\right)_M = -\frac{\Phi_\lambda}{T}$ follows from the constant-mass constraint dM = 0. Carrying out the differentiation explicitly gives the closed form $\mu_{JT} = -\frac{8\pi^4 K^2 \lambda}{\sqrt{\pi G}\, S^{3/2}(S(1-8\pi GK)-4\pi^2 K\lambda)}$. In the physical region the denominator is positive (T > 0 requires $S(1-8\pi GK) > 4\pi^2 K\lambda$), so for K, λ > 0 the coefficient is manifestly negative. Hence $\mu_{JT} < 0$ everywhere: the black hole always cools as λ increases at fixed mass, the equation $\mu_{JT} = 0$ has no solution, and no inversion temperature exists.

### 4.2 Results

Figure 5 shows the isenthalpic curves T(λ) at fixed black hole mass M for K = 0.015. Every curve decreases monotonically with λ, in agreement with the analytic result $\mu_{JT} < 0$ obtained above: at fixed mass the temperature falls as the quartic coupling grows. Because $\mu_{JT}$ never vanishes in the physical domain, there is no inversion temperature, and the entire physical region forms a single cooling regime—in contrast to charged AdS black holes, where an inversion curve separates heating and cooling.

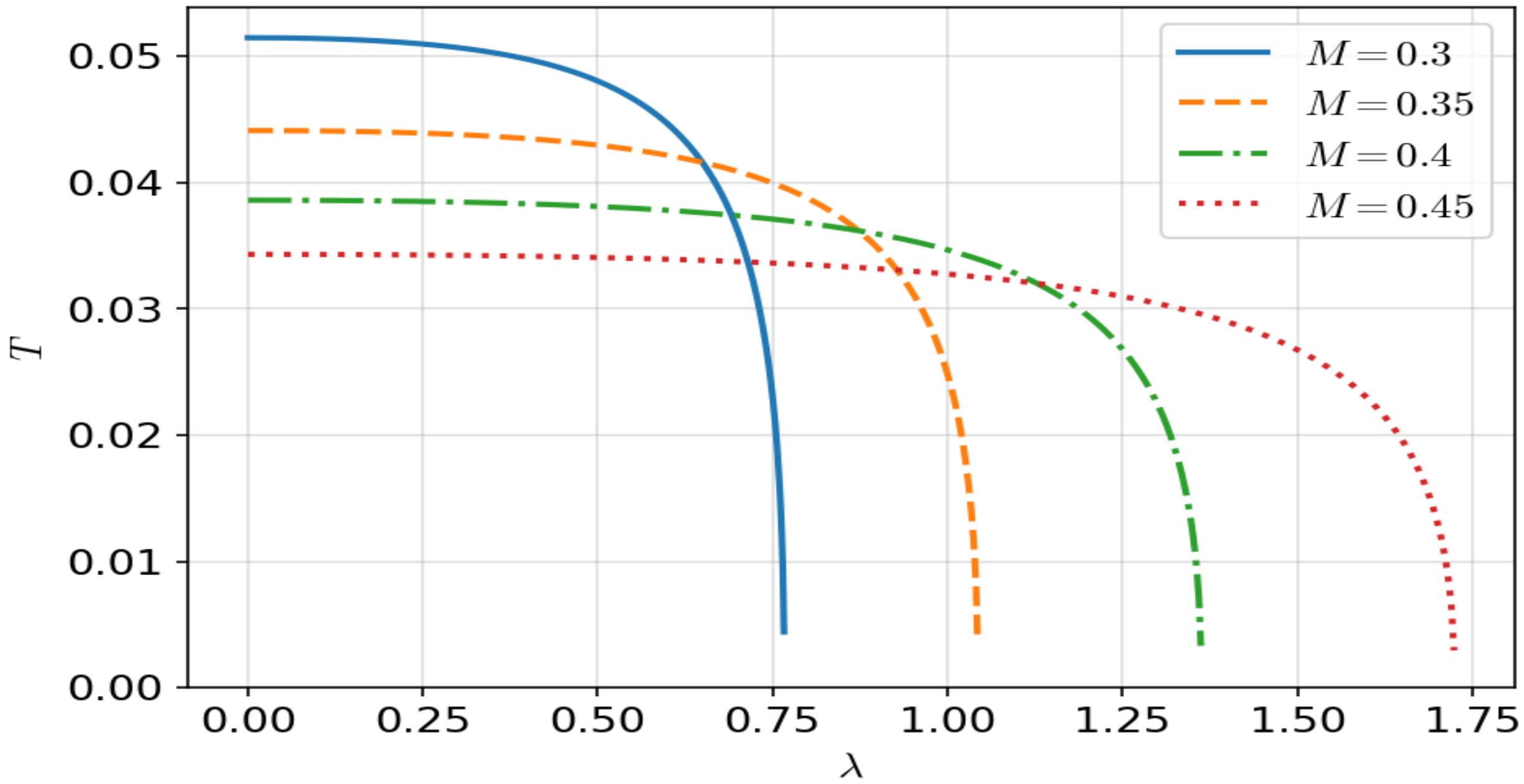

**Fig. 5** Isenthalpic curves T(λ) at fixed black hole mass M for K = 0.015 and G = 1. All curves are monotonically decreasing ($\mu_{JT} < 0$ everywhere), indicating that the Einstein–Skyrme black hole always cools upon increasing λ at constant M. Each curve terminates where the black hole becomes extremal (T → 0); the monotonic decrease all the way to extremality, with no turning point, is the graphical counterpart of $\mu_{JT} < 0$ everywhere and the absence of an inversion temperature.

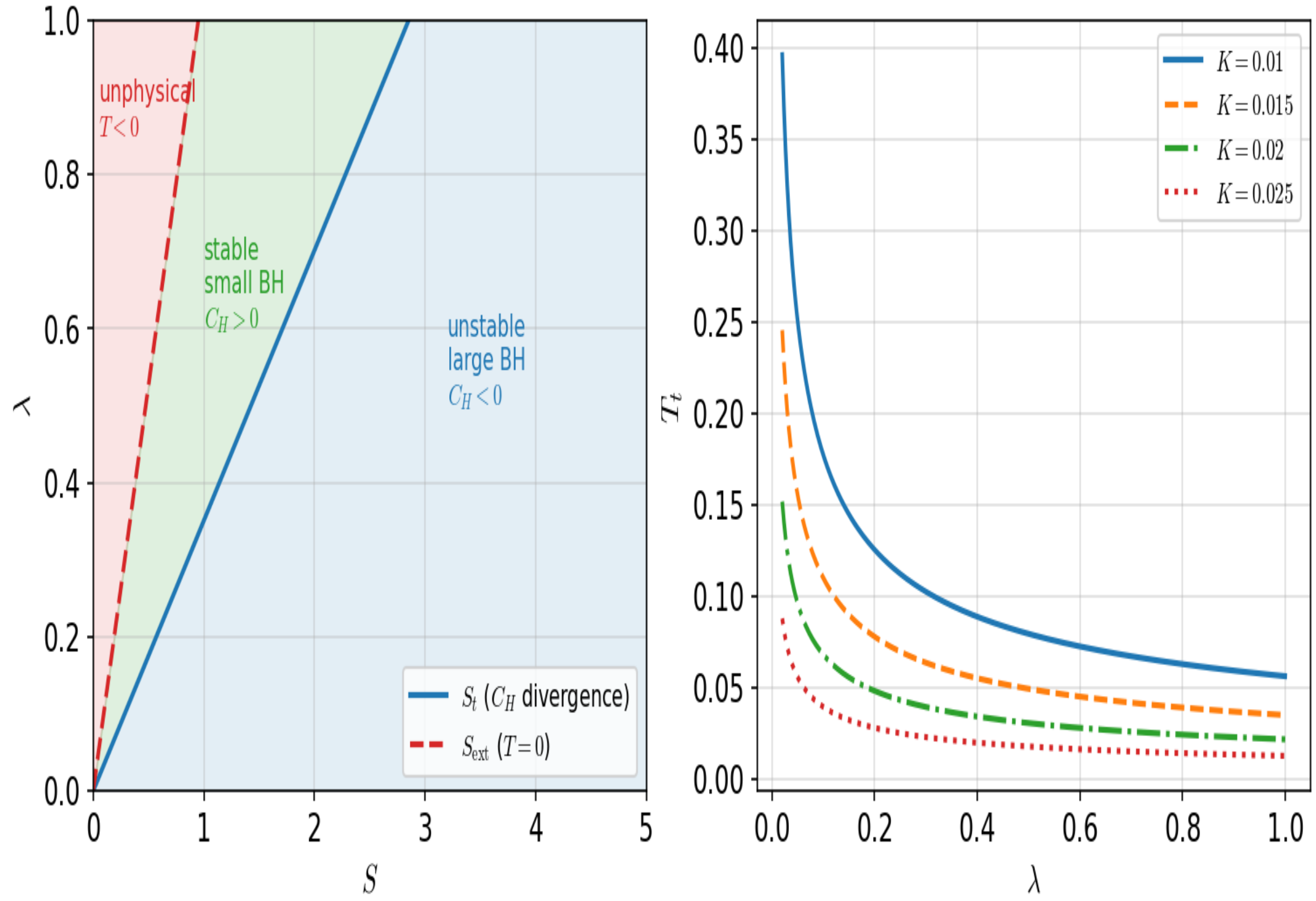

**Fig. 6** Left: diagram in the S–λ plane for K = 0.015. The extremal line S_ext (T = 0, dashed red) and the heat-capacity transition line $S_t = 3S_{ext}$ (solid blue) bound the unphysical region (T < 0), the stable small–black-hole branch ($C_H > 0$), and the unstable large–black-hole branch ($C_H < 0$). Right: transition temperature $T_t(\lambda)$for various K, showing the $T_t \propto \lambda^{-1/2}$ decrease with the quartic Skyrme coupling.

Figure 6 presents this diagram (left panel) and the transition temperature as a function of λ (right panel). The right panel confirms that $T_t$ decreases monotonically with λ; explicitly, since $S_t = \frac{12\pi^2 K\lambda}{1-8\pi GK} \propto \lambda$, substitution into Eq. (5) gives $T_t = \frac{1-8\pi GK}{6\sqrt{\pi G\, S_t}} \propto \lambda^{-1/2}$ at fixed K. A key result of the isenthalpic analysis is that the Joule–Thomson coefficient is strictly negative throughout the entire physical parameter space. Unlike charged AdS black holes—which possess a well-defined inversion temperature separating heating and cooling regimes [43–45]—the Einstein–Skyrme black hole always cools when λ increases at fixed mass. This absence of a Joule–Thomson inversion is a distinctive signature of the Skyrme matter sector: the quartic coupling λ enhances the effective charge-like contribution in the metric, systematically lowering the Hawking temperature, so that the entire physical domain forms a single cooling region. In the left panel the transition line $S_t = \left[\frac{12\pi^2 K}{1-8\pi GK}\right]\lambda$ is plotted with S on the horizontal axis and λ on the vertical axis, so its displayed slope is $\frac{d\lambda}{dS} = \frac{1-8\pi GK}{12\pi^2 K} \approx 0.35$ for K = 0.015.

## 5 Scalar perturbations, effective potential, and greybody factors

Having analyzed the thermodynamic microstructure, we now turn to the perturbative sector of the Einstein–Skyrme black hole. In this section, we derive the effective potential governing massless scalar field perturbations and establish rigorous bounds on the associated greybody factors.

### 5.1 Scalar field equation and effective potential

Consider a massless scalar field Φ propagating on the Einstein–Skyrme background (1)–(2). The Klein–Gordon equation Φ = 0 admits the mode decomposition: Because the background is spherically symmetric, the radial equation is independent of the azimuthal number m, so we write $\Psi_l$ rather than $\Psi_{lm}$ in what follows.

$$\Phi(t, r, \vartheta, \varphi) = \sum_{l,m} \frac{1}{r} \Psi_l(r) Y_{lm}(\vartheta, \varphi) e^{-i\omega t} \tag{14}$$

where $Y_{lm}$ are spherical harmonics. Introducing the tortoise coordinate r* defined by $\frac{dr^*}{dr} = \frac{1}{f(r)}$, the radial equation reduces to a Schrödinger-like form:

$$\frac{d^2\Psi_l}{dr^{*2}} + [\omega^2 - V_{eff}(r)]\, \Psi_l = 0 \tag{15}$$

with the effective potential:

$$V_{eff}(r) = f(r)\, [\frac{l(l+1)}{r^2} + \frac{f\prime(r)}{r}] \tag{16}$$

Substituting the metric function (2), we obtain:

$$V_{eff}(r) = \left(1 - 8\pi GK - \frac{2GM}{r} + \frac{4\pi GK\lambda}{r^2}\right) \left[\frac{l(l+1)}{r^2} + \frac{1}{r}\left(\frac{2GM}{r^2} - \frac{8\pi GK\lambda}{r^3}\right)\right] \tag{17}$$

The Skyrme parameters enter the effective potential through two distinct channels. The coupling K modifies both the constant and the $r^{-2}$ pieces of f(r), altering the overall height and asymptotic behavior of $V_{eff}$. The quartic coupling λ contributes additional $r^{-2}$ and higher-order terms, shifting the peak position and width of the potential barrier.

## 5.2 Effect of Skyrme couplings on the potential barrier

Figure 7 displays the effective potential for the l = 2 scalar mode as a function of the radial coordinate r, for fixed K = 0.015 and several values of λ. In the Schwarzschild limit (K = λ = 0) one recovers the standard Regge–Wheeler potential, whose peak for M = 1 is $V_0 \approx 0.247$. Switching on the solid-angle-deficit coupling K strongly lowers the barrier: already at K = 0.015 the peak falls to $V_0 \approx 0.058$ and moves outward (the horizon grows from $r_H = 2$ to $r_H \approx 3.2$). By contrast, at fixed K the quartic coupling λ has only a small effect: raising λ from 0 to 0.5 increases the peak by little more than one percent, consistent with the near-overlap of the curves in Fig. 7. This is expected, since the term $\frac{4\pi GK\lambda}{r^2}$ enters f(r) with a positive sign, exactly like the charge term $Q^2/r^2$ of a Reissner–Nordström black hole; it therefore acts repulsively, slightly raising rather than deepening the barrier. The dominant control parameter for the perturbative response is thus K, not λ.

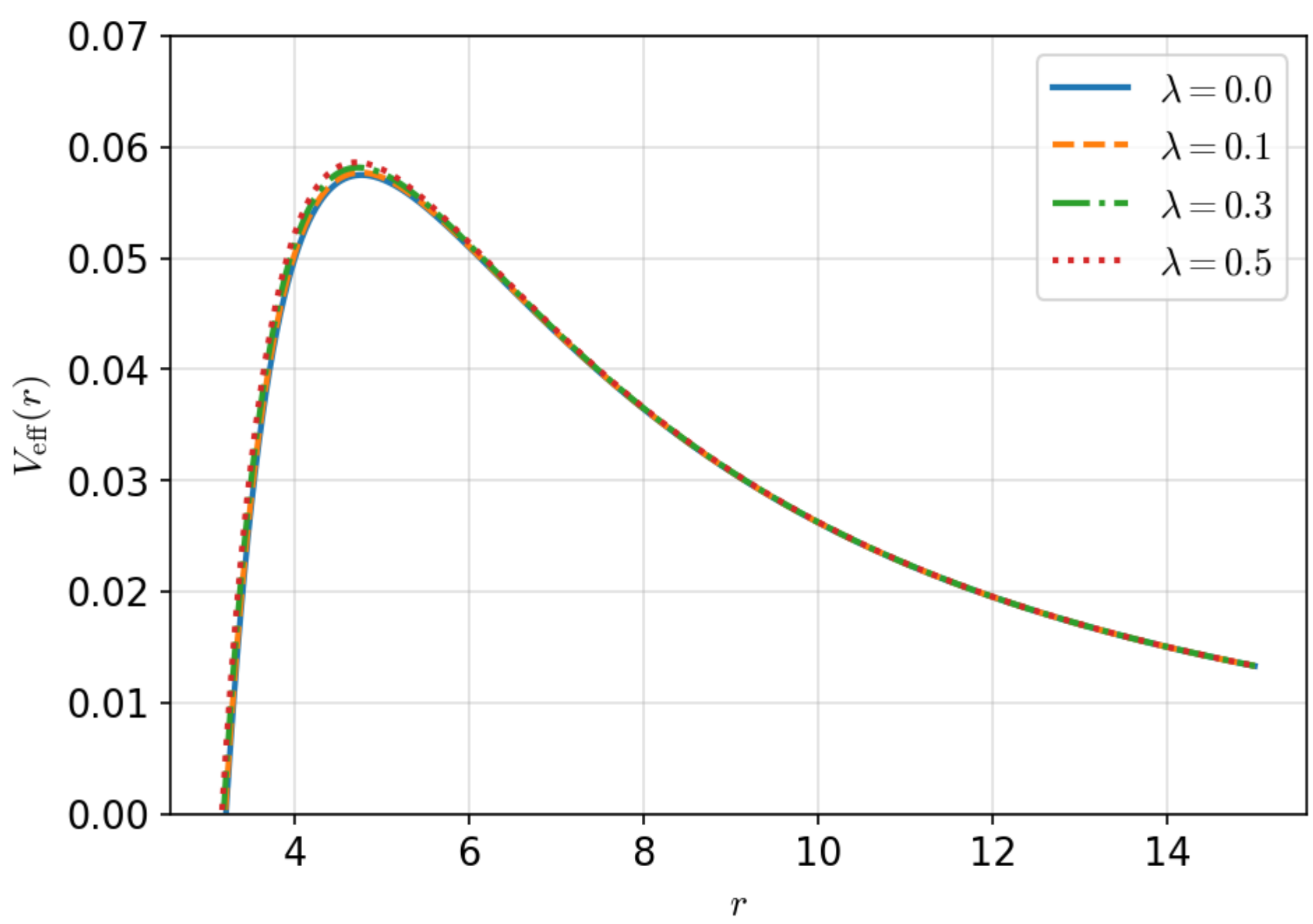

**Fig. 7** Effective potential $V_{eff}$ for scalar perturbations with l = 2 as a function of r for various λ, with K = 0.015, M = 1, and G = 1.

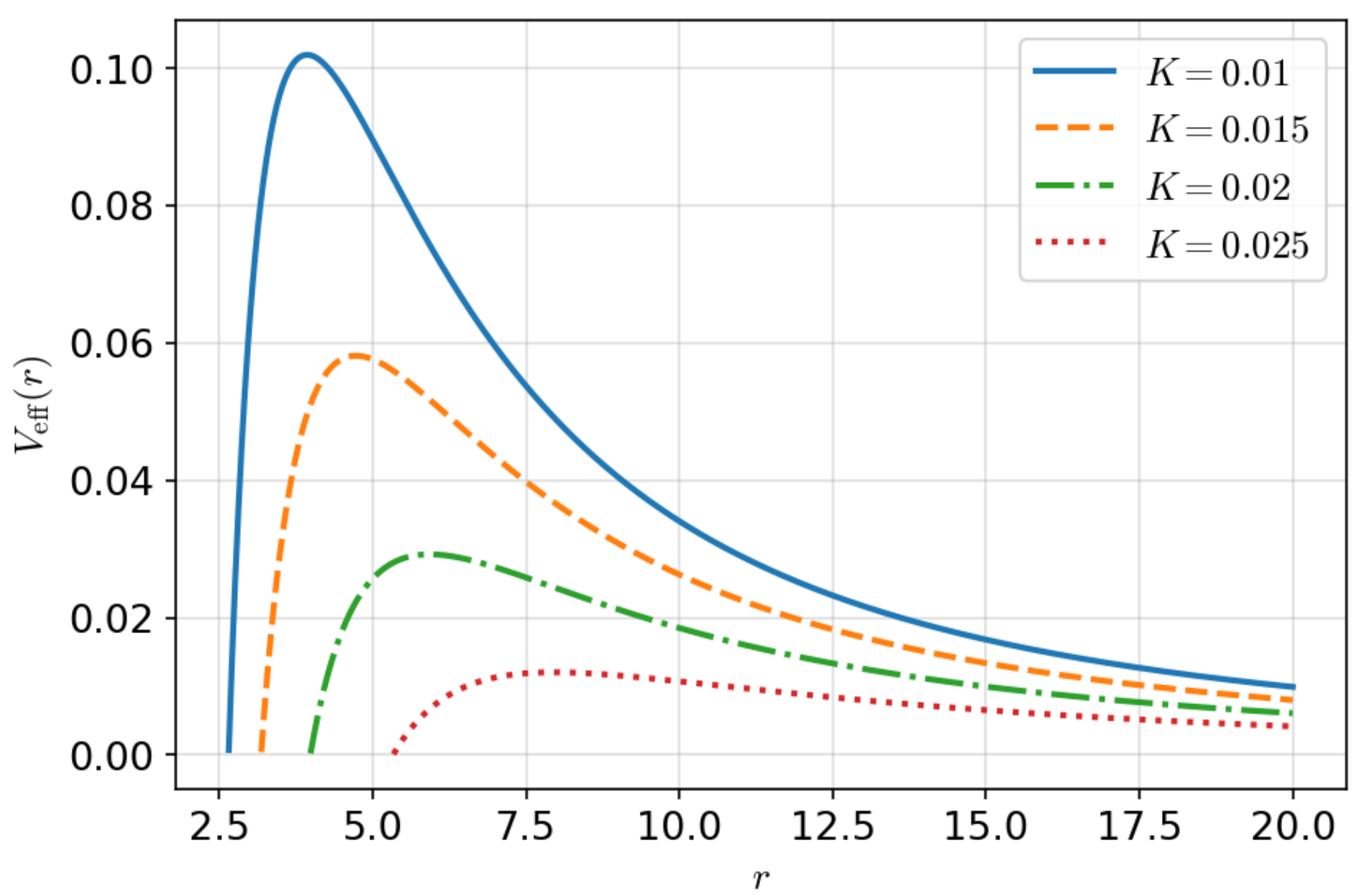

**Fig. 8** Effective potential $V_{eff}$ for scalar perturbations with l = 2 as a function of r for various K, with λ = 0.3, M = 1, and G = 1.

Figure 8 shows the effect of varying K at fixed λ = 0.3. Increasing K lowers the overall height of the potential, consistent with the interpretation that the solid-angle deficit reduces the effective gravitational potential barrier. This modification is expected to suppress the oscillation frequencies and enhance the damping rates of the quasinormal mode spectrum. We have verified that an event horizon exists for every parameter set shown (e.g. for K = 0.025, λ = 0.3 it lies at $r_H \approx 5.33$) and that $V_{eff} > 0$ throughout the exterior; the low, broad peak at K = 0.025 is physical, the larger horizon pushing the maximum outward to larger r.

### 5.3 WKB estimates of quasinormal frequencies

To obtain quantitative estimates we apply the WKB method [28–30, 60]. At leading (first) order the quasinormal frequencies are fixed by the height and curvature of the potential peak:

$$\omega^2 \approx V_0 - i\left(n + \frac{1}{2}\right)\sqrt{-2V_0''} \tag{18}$$

where $V_0 = V_{eff}(r_0)$ is the maximum of the potential at $r_0$ and $V_0'' = \left.\frac{d^2V_{eff}}{dr^2}\right|_{r_0}$ is its second derivative with respect to the tortoise coordinate, which at the peak equals $f(r_0)^2\, V_{eff}''(r_0)$. Table 2 lists the fundamental (n = 0) leading-order WKB frequencies for the l = 2 scalar mode at representative couplings. As a calibration, for Schwarzschild (M = 1) the leading-order formula gives ω ≈ 0.506 − 0.096i, about 5% above the accurate value ω ≈ 0.4836 − 0.0968i [30]; leading-order WKB is thus adequate for the qualitative trends reported here but not for high-precision frequencies, which would require the sixth-order WKB scheme with Padé resummation [60]. We verified numerically that $V_{eff}$ has a single maximum outside the horizon for every parameter set used, so that $V_0'' < 0$ and the square root in Eq. (18) is real.

**Table 2** Leading-order WKB quasinormal frequencies for the fundamental (n = 0, l = 2) scalar mode, with M = 1 and G = 1.

| K | λ | $\omega_r$ (oscillation) | $\lvert\omega_i\rvert$ (damping) | $V_0$ (peak) |
|---|---|---|---|---|
| 0.000 | 0.0 | 0.5063 | 0.0961 | 0.2471 |
| 0.015 | 0.0 | 0.2426 | 0.0373 | 0.0575 |
| 0.015 | 0.1 | 0.2431 | 0.0373 | 0.0577 |
| 0.015 | 0.3 | 0.2440 | 0.0374 | 0.0581 |
| 0.015 | 0.5 | 0.2450 | 0.0374 | 0.0586 |
| 0.025 | 0.3 | 0.1105 | 0.0133 | 0.0120 |

Table 2 shows that the quadratic coupling K is the dominant influence on the spectrum: increasing K markedly lowers V_0 and suppresses both ω_R and |ω_I|. At fixed K = 0.015, by contrast, varying λ over 0–0.5 changes the peak and the frequencies by only about one percent, with a slight increase rather than a decrease. The earlier expectation that λ strongly lowers the frequencies is therefore not borne out; the λ-dependence at fixed K is weak, in agreement with the near-coincident curves of Fig. 7.

## 5.4 Greybody factor bounds

The greybody factor $\mathcal{T}_l(\omega)$— distinct from the Hawking temperature T — quantifies the probability that a quantum of Hawking radiation with frequency ω is transmitted through the effective potential barrier surrounding the black hole. Rather than computing the exact transmission coefficient, we establish rigorous lower bounds using the general method of Visser [46]:

$$T_l(\omega) \geq \mathrm{sech}^2\left(\frac{1}{2\omega}\int_{-\infty}^{\infty}|V_{eff}(r^*)|dr^*\right) \tag{19}$$

For the present background the integral can be evaluated in closed form. Since $\frac{V_{eff}}{f} = \frac{l(l+1)}{r^2} + \frac{f'}{r}$, the tortoise integral collapses to $\int V_{eff}\, dr^* = \int_{r_H}^{\infty}\frac{V_{eff}}{f}\, dr = \frac{l(l+1)}{r_H} + \frac{GM}{r_H^2} - \frac{8\pi GK\lambda}{3r_H^3}$, with $r_H = \frac{1+\sqrt{1-4\pi GK\lambda(1-8\pi GK)}}{1-8\pi GK}$. The would-be solid-angle-deficit prefactor (1−8πGK) cancels between V_eff and the tortoise measure dr* = dr/f, so the bound is finite despite the conical asymptotics. The angular channels represent the relevant cases: l = 1 is the lowest multipole carrying a centrifugal barrier, a convenient representative for transmission, while the quasinormal analysis uses l = 2, the dominant multipole for a massless scalar. For completeness, the s-wave l = 0—which dominates the low-energy Hawking flux—obeys the same formula with l(l+1) = 0, giving $\int = \frac{GM}{r_H^2} - \frac{8\pi GK\lambda}{3r_H^3}$ and a correspondingly larger low-frequency transmission. Figure 9 shows the bounds for the l = 1 mode versus frequency ω for several λ at fixed K = 0.015.

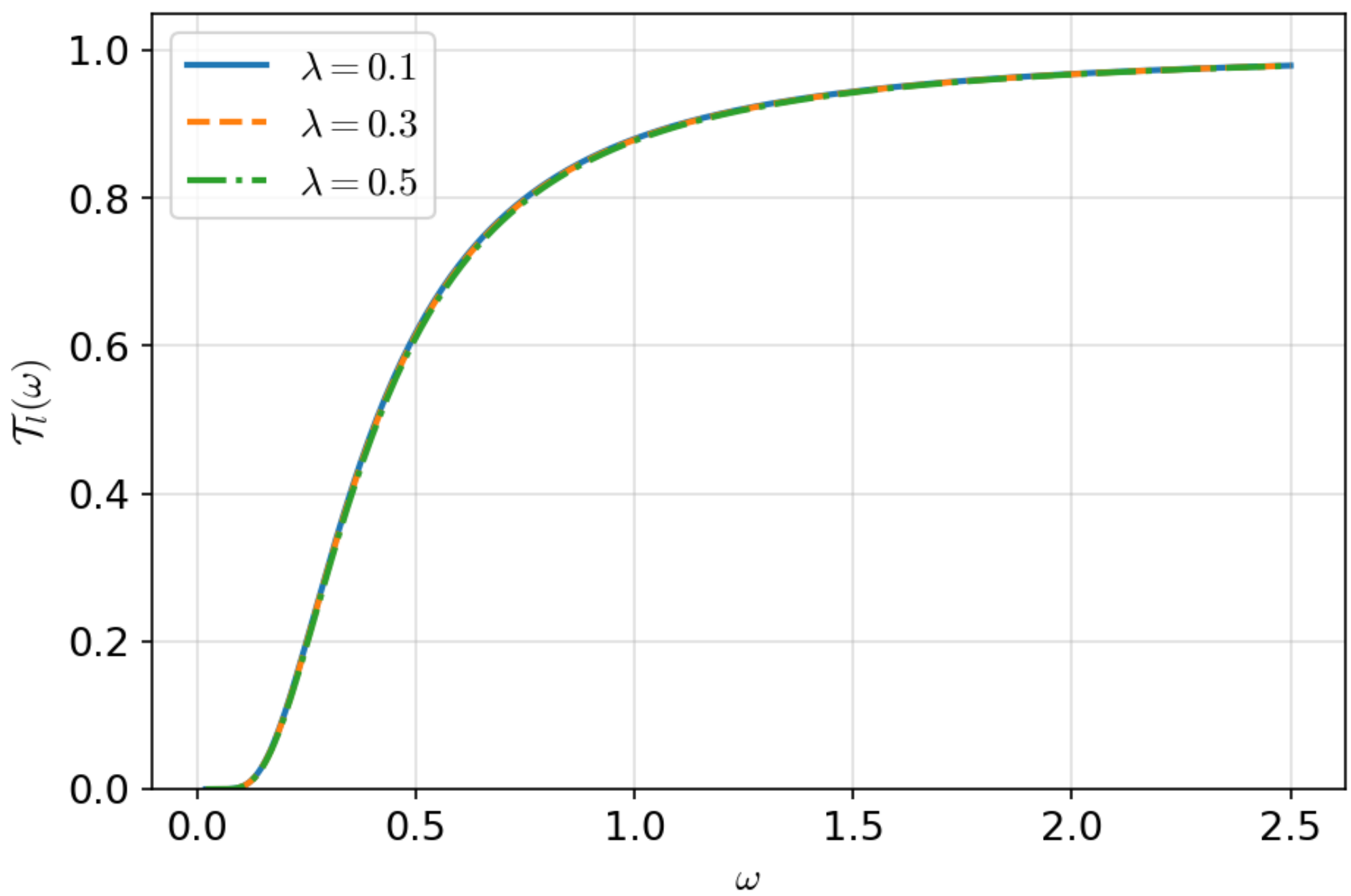

**Fig. 9** Rigorous lower bound on the greybody factor $T_l$ for the l = 1 scalar mode as a function of frequency ω for various λ, with K = 0.015, M = 1, and G = 1.

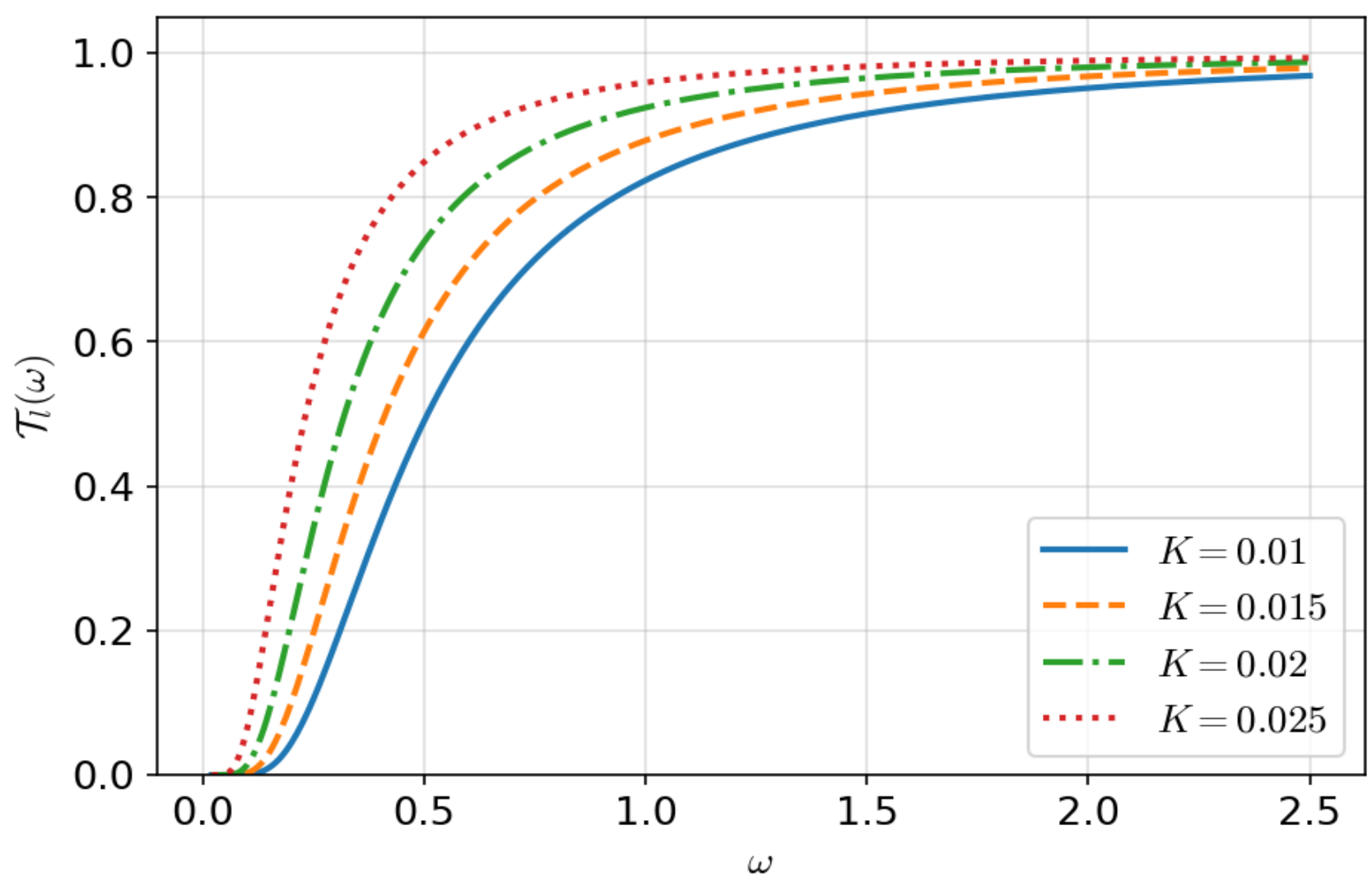

**Fig. 10** Rigorous lower bound on the greybody factor $T_l$ for the l = 1 scalar mode for various K, with λ = 0.3, M = 1, and G = 1.

Several features are evident from Figs. 9 and 10. The dominant trend is set by K: increasing the solid-angle-deficit coupling enlarges the horizon and lowers the centrifugal barrier $l(l+1)/r_H$, which reduces the tortoise integral and therefore raises the greybody bound—more low-frequency radiation is transmitted as K grows. At fixed K, increasing λ slightly reduces the low-frequency bound, but the effect is small (a few percent over 0 ≤ λ ≤ 0.5), consistent with the near-overlap of the curves in Fig. 9. At high frequencies ($\omega \gg V_0$) the bound approaches unity for all couplings, as expected for short-wavelength radiation. We caution that comparisons across different K mix two effects—the change in horizon size and the change in asymptotic (conical) structure—so the absolute transmission at fixed ω across different K should be read with the rescaled, asymptotically flat frame in mind.

**Table 3** Greybody factor bound $T_l$ at selected frequencies for the l = 1 scalar mode (M = 1, G = 1)

| K | λ | $T_l$(ω=0.2) | $T_l$(ω=0.5) | $T_l$(ω=1.0) | $T_l$(ω=2.0) |
|---|---|---|---|---|---|
| 0.000 | 0.0 | 0.008 | 0.280 | 0.692 | 0.908 |
| 0.015 | 0.1 | 0.103 | 0.618 | 0.880 | 0.968 |
| 0.015 | 0.3 | 0.101 | 0.615 | 0.879 | 0.968 |
| 0.015 | 0.5 | 0.099 | 0.612 | 0.878 | 0.967 |
| 0.025 | 0.3 | 0.405 | 0.849 | 0.959 | 0.990 |

Table 3 lists the bound at representative frequencies for the l = 1 mode. Across fixed K = 0.015, raising λ from 0.1 to 0.5 lowers the ω = 0.2 transmission only modestly, by roughly 4%. The stronger variation is with K: at ω = 0.2 the bound rises from about 0.01 in the Schwarzschild case to roughly 0.10 at K = 0.015 and 0.40 at K = 0.025, reflecting the growth of the horizon radius. The Skyrme sector thus reshapes the Hawking transmission spectrum, with the solid-angle-deficit coupling K—rather than the quartic coupling λ—playing the leading role.

# 6 Conclusions

In this work, we have provided the first comprehensive investigation of the thermodynamic microstructure, Joule–Thomson expansion, quasinormal mode spectrum, and greybody factors of the exact, static, spherically symmetric Einstein–Skyrme black hole, building upon the recently formulated extended first law of Senjaya and Klaypong [42]. Our main findings are as follows.

First, we constructed the genuine two-dimensional Ruppeiner thermodynamic geometry of the Einstein–Skyrme black hole in the extended phase space where K and λ are promoted to thermodynamic variables, obtaining the scalar curvature in the closed form $R = \frac{1}{S_{\text{ext}}-S}$, with extremal entropy $S_{\text{ext}} = \frac{4\pi^2 K\lambda}{1-8\pi GK}$. The curvature is finite at the second-order (heat-capacity) transition $S_t = 3S_{\text{ext}}$ and diverges instead in the extremal, zero-temperature limit $S \to S_{\text{ext}}$, where the small–black-hole branch terminates. Within the entire physical region $S > S_{\text{ext}}$ the curvature keeps a single sign (no repulsive–attractive crossover), indicating microscopic interactions of one dominant character that strengthens toward extremality and weakens for large black holes. The sign itself is convention dependent, so we have stated these conclusions in convention-independent terms. This absence of a sign change stands in contrast to the Wei–Liu–Mann sign-change picture established for charged AdS black holes [19, 25] and constitutes a distinctive result of the present analysis rather than a limitation of it. The curvature magnitude decreases monotonically with entropy, $|R| \to 0$ as $S \to \infty$, so large black holes approach non-interacting behaviour; the strict Schwarzschild limit K, λ → 0 is degenerate (the coupling directions drop out of the metric) and is therefore approached rather than taken.

Second, we performed a Joule–Thomson-like isenthalpic expansion analysis and constructed a diagram in the (S, λ) plane delineating the unphysical, stable small–black-hole, and unstable large–black-hole regions. We proved analytically that the Joule–Thomson coefficient is strictly negative throughout the physical parameter space—the black hole always cools when λ increases at fixed mass—so that no inversion temperature exists, distinguishing the Einstein–Skyrme system from charged AdS black holes. We also obtained the transition temperature $T_t(\lambda)$, which decreases as $T_t \propto \lambda^{-1/2}$ at fixed K. The distinctive thermodynamic character is driven by matter-sector couplings rather than by a cosmological constant.

Third, we derived the effective potential for massless scalar perturbations on the Einstein–Skyrme background and found that the solid-angle-deficit coupling K dominates its shape, strongly lowering the barrier, whereas at fixed K the quartic coupling λ produces only a small (slightly raising) change—since λ enters f(r) with the same positive sign as a Reissner–Nordström charge term. Using leading-order WKB estimates (calibrated against the known Schwarzschild value and supplemented by the higher-order scheme of Ref. [60] where precision is required), we showed that the fundamental quasinormal frequencies decrease with K, while the λ-dependence at fixed K is weak. Rigorous closed-form lower bounds on the greybody factors were obtained by Visser's method: increasing K raises the bound, because the larger horizon lowers the centrifugal barrier, while increasing λ at fixed K mildly suppresses the low-frequency transmission. The Skyrme sector thus reshapes the Hawking emission spectrum, with K playing the leading role. We emphasize that this dominance of the solid-angle-deficit coupling K over the quartic coupling λ in the perturbative sector is itself a non-trivial finding: although λ enters the lapse function exactly like a Reissner–Nordström charge, the Einstein–Skyrme black hole does not behave as a naive Reissner–Nordström analogue—its quasinormal spectrum and Hawking transmission are governed primarily by the global conical structure of the geometry rather than by the effective charge.

Our results open several avenues for future investigation. The extension of the Ruppeiner analysis to the full two-dimensional thermodynamic space (S, λ) would require finding a non-degenerate metric, possibly through the inclusion of additional thermodynamic variables or by considering the Einstein–Skyrme system with a cosmological constant. The computation of exact quasinormal frequencies using higher-order WKB or spectral methods, and the inclusion of electromagnetic and gravitational perturbations, would provide a more complete perturbative characterization. The connection between the microstructural interactions identified here and the topological properties of the underlying Skyrmion field is another direction worthy of exploration. Finally, applying the present framework to the pion-sourced black holes recently discussed in [47] would extend the analysis to non-spherically symmetric hadronic configurations.

## References

[1] J.D. Bekenstein, Black holes and the second law, Lett. Nuovo Cim. 4, 737 (1972).

[2] J.D. Bekenstein, Black holes and entropy, Phys. Rev. D 7, 2333 (1973).

[3] S.W. Hawking, Black Holes and Thermodynamics, Phys. Rev. D 13, 191 (1976).

[4] W. Israel, Event horizons in static vacuum space-times, Phys. Rev. 164, 1776 (1967).

[5] B. Carter, Axisymmetric Black Hole Has Only Two Degrees of Freedom, Phys. Rev. Lett. 26, 331 (1971).

[6] R.N. Izmailov, E.R. Zhdanov, A. Bhadra, K.K. Nandi, Relative time delay in a spinning black hole, Eur. Phys. J. C 79, 105 (2019).

[7] Y.-L. Ma, M. Rho, Dichotomy of Baryons as Quantum Hall Droplets and Skyrmions, Symmetry 13, 1888 (2021).

[8] B.P. Dolan, The cosmological constant and the black hole equation of state, Class. Quant. Grav. 28, 125020 (2011).

[9] D. Kubiznak, R.B. Mann, M. Teo, Black hole chemistry: thermodynamics with Lambda, Class. Quant. Grav. 34, 063001 (2017).

[10] R.B. Mann, Recent Developments in Holographic Black Hole Chemistry, JHAP 4, 1 (2024).

[11] A. Chamblin, R. Emparan, C.V. Johnson, R.C. Myers, Holography, thermodynamics and fluctuations of charged AdS black holes, Phys. Rev. D 60, 104026 (1999).

[12] D. Kubiznak, R.B. Mann, P-V criticality of charged AdS black holes, JHEP 07, 033 (2012).

[13] S.-W. Wei, P. Cheng, Y.-X. Liu, Analytical critical phenomena of singly spinning Kerr-AdS black holes, Phys. Rev. D 93, 084015 (2016).
[14] P. Cheng, S.-W. Wei, Y.-X. Liu, Critical phenomena in Kerr-Newman-AdS black holes, Phys. Rev. D 94, 024025 (2016).
[15] R.A. Hennigar, R.B. Mann, E. Tjoa, Superfluid Black Holes, Phys. Rev. Lett. 118, 021301 (2017).
[16] A.M. Frassino, D. Kubiznak, R.B. Mann, F. Simovic, Multiple Reentrant Phase Transitions and Triple Points in Lovelock Thermodynamics, JHEP 09, 080 (2014).
[17] M. Tavakoli, J. Wu, R.B. Mann, Multi-critical points in black hole phase transitions, JHEP 12, 117 (2022).
[18] A.M. Frassino, J.F. Pedraza, A. Svesko, M.R. Visser, Higher-Dimensional Origin of Extended Black Hole Thermodynamics, Phys. Rev. Lett. 130, 161501 (2023).
[19] S.-W. Wei, Y.-X. Liu, R.B. Mann, Repulsive Interactions and Universal Properties of Charged Anti-de Sitter Black Hole Microstructures, Phys. Rev. Lett. 123, 071103 (2019).
[20] G. Ruppeiner, Riemannian geometry in thermodynamic fluctuation theory, Rev. Mod. Phys. 67, 605 (1995).
[21] G. Ruppeiner, Thermodynamic curvature and black holes, Springer Proc. Phys. 153, 179 (2014).
[22] G. Ruppeiner, Thermodynamic curvature and phase transitions in Kerr-Newman black holes, Phys. Rev. D 78, 024016 (2008).
[23] Z.-M. Xu, B. Wu, W.-L. Yang, Ruppeiner thermodynamic geometry for the Schwarzschild-AdS black hole, Phys. Rev. D 101, 024018 (2020).
[24] A. Ghosh, C. Bhamidipati, Thermodynamic geometry and interacting microstructures of BTZ black holes, Phys. Rev. D 101, 106007 (2020).
[25] S.-W. Wei, Y.-X. Liu, R.B. Mann, Ruppeiner Geometry, Phase Transitions, and the Microstructure of Charged AdS Black Holes, Phys. Rev. D 100, 124033 (2019).
[26] X.-Y. Guo, H.-F. Li, L.-C. Zhang, R. Zhao, Microstructure and continuous phase transition of a Reissner-Nordstrom-AdS black hole, Phys. Rev. D 100, 064036 (2019).
[27] S.-J. Yang, R. Zhou, S.-W. Wei, Y.-X. Liu, Ruppeiner geometry and thermodynamic phase transition of black hole in massive gravity, Eur. Phys. J. C 81, 545 (2021).
[28] B.F. Schutz, C.M. Will, Black hole normal modes: a semianalytic approach, Astrophys. J. 291, L33 (1985).
[29] S. Iyer, C.M. Will, Black-hole normal modes: A WKB approach, Phys. Rev. D 35, 3621 (1987).
[30] R.A. Konoplya, A. Zhidenko, Quasinormal modes of black holes: from astrophysics to string theory, Rev. Mod. Phys. 83, 793 (2011).
[31] P. Kanti, J. March-Russell, Calculable corrections to brane black hole decay, Phys. Rev. D 66, 024023 (2002).
[32] R.A. Konoplya, A.F. Zinhailo, Grey-body factors and Hawking radiation of black holes in 4D Einstein-Gauss-Bonnet gravity, Phys. Lett. B 810, 135793 (2020).
[33] T.H.R. Skyrme, A Unified Field Theory of Mesons and Baryons, Nucl. Phys. 31, 556 (1962).
[34] G.S. Adkins, C.R. Nappi, E. Witten, Static Properties of Nucleons in the Skyrme Model, Nucl. Phys. B 228, 552 (1983).
[35] F. Canfora, H. Maeda, Hedgehog ansatz and its generalization for self-gravitating Skyrmions, Phys. Rev. D 87, 084049 (2013).
[36] N. Shiiki, N. Sawado, Regular and black hole solutions in the Einstein-Skyrme theory with negative cosmological constant, Class. Quant. Grav. 22, 3561 (2005).

[37] Y. Brihaye, T. Delsate, Skyrmion and Skyrme-black holes in de Sitter spacetime, Mod. Phys. Lett. A 21, 2043 (2006).
[38] A.B. Nielsen, Skyrme Black Holes in the Isolated Horizons Formalism, Phys. Rev. D 74, 044038 (2006).
[39] M. Barriola, A. Vilenkin, Gravitational Field of a Global Monopole, Phys. Rev. Lett. 63, 341 (1989).
[40] D. Flores-Alfonso, H. Quevedo, Extended Thermodynamics of Self-Gravitating Skyrmions, Class. Quant. Grav. 36, 154001 (2019).
[41] Y.H. Khan, P.A. Ganai, S. Upadhyay, Quantum-corrected thermodynamics and P–V criticality of self-gravitating Skyrmion black holes, PTEP 2020, 103B06 (2020).
[42] D. Senjaya, A. Klaypong, Einstein–Skyrme black hole's novel first law of Skyrmion's black hole thermodynamics, Eur. Phys. J. C 86, 309 (2026).
[43] Ö. Ökcü, E. Aydıner, Joule–Thomson expansion of the charged AdS black holes, Eur. Phys. J. C 77, 24 (2017).
[44] M. Chabab, H. El Moumni, S. Iresi, K. Masmar, Joule-Thomson Expansion of RN-AdS Black Holes in f(R) gravity, Lett. High Energy Phys. 02, 05 (2018).
[45] C.L.A. Rizwan, A.N. Kumara, D. Vaid, K.M. Ajith, Joule–Thomson expansion in AdS black hole with a global monopole, Int. J. Mod. Phys. A 33, 1850210 (2019).
[46] M. Visser, Some general bounds for one-dimensional scattering, Phys. Rev. A 59, 427 (1999).
[47] F. Canfora, A. Gomberoff, C. Henríquez-Baez, A. Vera, Exact black holes with bumpy horizons supported by superfluid pions, arXiv:2601.22914 [hep-th] (2026).
[48] F. Canfora, E.F. Eiroa, C.M. Sendra, Spherical skyrmion black holes as gravitational lenses, Eur. Phys. J. C 78, 659 (2018).
[49] G.W. Gibbons, Causality and the Skyrme model, Phys. Lett. B 566, 171 (2003).
[50] F. Canfora, F. Correa, J. Zanelli, Exact multisoliton solutions in the four-dimensional Skyrme model, Phys. Rev. D 90, 085002 (2014).
[51] K. Hajian, B. Tekin, Coupling Constants as Conserved Charges in Black Hole Thermodynamics, Phys. Rev. Lett. 132, 191401 (2024).
[52] Y. Xiao, Y. Tian, Y.-X. Liu, Extended Black Hole Thermodynamics from Extended Iyer-Wald Formalism, Phys. Rev. Lett. 132, 021401 (2024).
[53] S.-J. Ma, R.-B. Wang, J.-B. Deng, X.-R. Hu, Euler-Heisenberg black hole surrounded by perfect fluid dark matter, Eur. Phys. J. C 84, 595 (2024).
[54] D. Senjaya, First law and black hole thermodynamics of black hole in Dehnen dark matter environment, Nucl. Phys. B 1024, 117341 (2026).
[55] A. Baruah, P. Phukon, CFT phase transition analysis of charged, rotating black holes, Phys. Rev. D 111, 066022 (2025).
[56] B.P. Dolan, Compressibility of rotating black holes, Phys. Rev. D 84, 127503 (2011).
[57] R.B. Mann, Black hole chemistry: The first 15 years, Int. J. Mod. Phys. D 34, 2542001 (2025).
[58] D.J. Gogoi, A. Övgün, D. Demir, Quasinormal modes and greybody factors of symmergent black hole, Phys. Dark Univ. 42, 101314 (2023).
[59] E. Khalaf, A. Vishwanath, Baby skyrmions in Chern ferromagnets, Nature Commun. 13, 6245 (2022).
[60] R.A. Konoplya, A. Zhidenko, A.F. Zinhailo, Higher order WKB formula for quasinormal modes and grey-body factors, Class. Quant. Grav. 36, 155002 (2019).